\documentstyle[aps]{revtex}
\begin{document} 
\vskip 2truecm
\title{Diffusion on non exactly decimable tree-like fractals}

\author{Raffaella Burioni\footnote{E-mail:
burioni@almite.mi.infn.it}}

\address{Dipartimento di Fisica,
Universit\`a di Milano, via Celoria 16, 20133 Milano, Italy}
\author{Davide Cassi\footnote{E-mail: cassi@vaxpr.pr.infn.it}, Alberto Pirati,
 Sofia 
Regina\footnote{E-mail: regina@vaxpr.pr.infn.it}  }
\address{ Dipartimento di Fisica, Universit\`a di Parma,
          Viale delle Scienze, 43100 Parma, Italy }
              
\maketitle
\vskip 1truecm
\begin{abstract} 

We calculate the spectral dimension of a wide class of tree-like
fractals by solving the random walk problem through a new analytical 
technique, based on invariance under generalized cutting-decimation 
transformations. 
These fractals are generalizations of the $NT_D$ lattices and they are
characterized by non integer spectral dimension equal or 
greater then 2, non anomalous diffusion laws, dynamical dimension splitting 
and absence of phase transitions for spin models.
\end{abstract}
\vskip 1.5truecm
UPRF-97-10
\newpage
\section{Introduction}
The spectral dimension $\widetilde{d}$ of non translation invariant structures
is up to now the best generalization of Euclidean dimension of lattices when
dealing with dynamical and thermodynamical properties. It can be defined
according to the large time asymptotic behavior of random walks and can be
shown to be relevant for several different physical phenomena, such as
vibrational dynamics, electrical conductivity and phase transitions.
Many properties of $\widetilde{d}$ has been argued and proved starting from 
exact analytical calculations on a few particular cases and any further
progress in understanding its relevance strongly depends on the availability of
exact results in a wider range of particular cases.
The most used mathematical tools to calculate $\widetilde{d}$ belong to
two main classes: renormalization group and combinatorial techniques.
Unfortunately renormalization group can give exact results only on exactly
decimable fractals which, in turn, have been shown to have $\widetilde{d}<2$.
On the other hand combinatorial techniques, based on iteration of cuttings, 
can be applied only to discrete
structures with a given characteristic scale. 
Recently, a particular combination of both techniques has been successfully
applied to the more 
complex case of $NT_D$ lattices \cite{cinque}.
These lattices are not exactly decimable but they are invariant 
under a more complex geometrical transformation we shall call 
Cutting-Decimation. 
On $NT_D$ lattices the random walks problem has been analytically solved using
a Cutting-Decimation method based on a time-rescaling technique \cite{sei}.
The $NT_D$ has been shown to have remarkable properties such as
non integer spectral dimension equal or greater then 2, non anomalous 
diffusion laws, absence of phase transitions \cite{sette} and dynamical 
dimension splitting \cite{otto}. 
Because of these peculiar characteristics $NT_D$ lattices have been widely 
used for the study of statistical mechanics on non translation invariant
lattices and they opened the way to the research of other structures sharing 
the same properties. 

In this paper we deal with generalizations of $NT_D$ lattices we shall call
$2^mNT_D$, $nNT_D$ and $p-polygon NT_D$ introducing three corresponding new 
techniques to solve analytically the random walks problem. 
As a result we obtain very general structures with the same properties of 
$NT_D$ and, at the same time, we test new techniques that can be useful
in other cases of non exactly decimable fractals. 

\section{Random walks on $2^m NT_D$ via exact time rescaling}
The $2^mNT_D$ are infinite fractal trees that can be recursively built using 
the following recipe (Fig. 1). 
An origin point $O$ is connected to point $A$ by a link 
of length 1 (the log of the tree); from $A$ the log splits in $k$ branches of 
length $2^m$ (i.e. made of $2^m$ consecutive links)
which, in turn, split in $k$ branches of length $2^{2m}$  
and so on in such a way that each branch of length $2^{nm}$ 
splits in $k$ branches of length $2^{(n+1)m}$. 
The simplest case is that of $m=1$ and corresponds to the $NT_D$ lattices. 
The $2^mNT_D$ invariance under the Cutting-Decimation transform can be 
described as follows
(Fig. 2). Suppose to cut the log of the tree and to separate the 
$k$ branches starting from $A$. Then each of these 
branches is exactly the same as the original tree with a dilatation 
factor equal to $2^m$ and therefore it can be reduced to the 
original tree through decimation. 
Let us now consider the discrete time 
random walks problem on $2^mNT_D$, in order to calculate the spectral dimension 
$\widetilde{d}$ of the lattice.
The spectral dimension of a graph is defined through the relation \cite{zero}:
\begin{equation}
P_O(t) \sim t^{- \widetilde{d} /2}
\end{equation}
where $P_O(t)$ is the probability for a random walker to return to the starting 
point $O$ after a walk of $t$ steps for $t\rightarrow \infty$. 
The cutting transformation applied to random walks gives 
a relation between $P_O^{\rm tree}(t)$, the probability of returning to the 
starting site $O$ after a $t$ steps walk on the whole tree, and $P_A^{\rm
branch}(t)$, the probability of returning to the starting site $A$ after a $t$
steps walks on one of the branches starting from $A$.  
This relation has been obtained in 
\cite{sei} for the case $m=1$ in terms of the generating
functions (discrete Laplace Transforms)
$\widetilde{P}(\lambda)$ of the probabilities $P(t)$, but it holds
for $2^mNT_D$ for every $m$:
\begin{equation}
\widetilde{P_O}^{tree}(\lambda)= {{\widetilde{P_A}^{branch}(\lambda)+k}
\over {(1-\lambda^2) {P_A}^{branch}(\lambda) +k }}
\label{halley}
\end{equation}
In the $m=1$ case the decimation transformation has been performed using a 
time-rescaling technique. Indeed, the motion of the 
random walker on the branch considered only after an even number of steps
can be exactly mapped in the motion of a random walker on the tree after 
the introduction of a staying probability $p_{ii}=1/2$ in every site $i$. 
This equivalence can be translated in terms of generating functions through the 
substitutions \cite{sei}:
 \begin{equation}
  \widetilde{P_O}(\lambda)\rightarrow {\lambda \over {2 -\lambda}}
\widetilde{P_O}\left(  {2 \over {2 -\lambda}} \right)
\label{uno}
\end{equation}
\begin{equation}
 \lambda \rightarrow  \lambda^2
\label{due}
\end{equation}
In the case $m=1$ equations (\ref{uno}) and (\ref{due}) can be used to rewrite 
(\ref{halley}) as:
\begin{equation}
\widetilde{P_O}^{tree}(\lambda)={{
{2 \over {2 -\lambda ^2}}\widetilde{P_O}^{tree}\left(
{\lambda ^2 \over { 2- \lambda ^2}}\right) +k} \over {
( 1- \lambda^2) {2 \over {2 -\lambda ^2}} \widetilde{P_O}^{tree}\left(
{\lambda ^2 \over { 2- \lambda ^2}}\right) +k}}
\end{equation}
If $m>1$ this procedure must be iterated $m$ times obtaining:
\begin{equation}
\widetilde{P_O}^{tree}(\lambda)={{
\left(
\prod_{i=1} ^{m} {2 \over {2 -\lambda_i ^2}} \right)
\widetilde{P_O}^{tree}(\lambda
_{i+1})+k} \over {(1-\lambda^2)\left( \prod_{i=1} ^{m} {2 \over {2 -\lambda_i 
^2}}  \right)\widetilde{P_O}^{tree}(\lambda _{i+1})+k}}
\label{muir}
\end{equation}
with
\[
\lambda_i = \cases  {\lambda& $i=1$ \cr
                      {} \cr
                     {{\lambda_{i-1}^2}\over {2- \lambda_{i-1}^2}}
  & $i>1$  \cr}
\]
$i$ being the iteration step.
To find the value of $\widetilde{d}$ we consider the singularities of the 
generating functions as $\lambda=1-\epsilon$, $\epsilon \rightarrow 0^+$.
In terms of $\epsilon$ the decimation transformation can be resumed as:
\begin{equation}
\widetilde{P_A}^{branch}(\epsilon)\sim 2^m \widetilde{P_O}^{tree}
(2^{2m}\epsilon)
 \end{equation}
as $\epsilon \rightarrow 0$, so that (\ref{muir}) becomes:
\begin{equation}
\widetilde{P_O}^{tree}(\epsilon) \sim {{ 2^m \widetilde{P_O}^{tree}
\left( 2^{2m} \epsilon \right) +k} \over { 2\epsilon 2^m \widetilde{P_O}^{tree}
\left( 2^{2m} \epsilon \right)+k}} 
\label{rel}
\end{equation}
From (\ref{rel}), applying standard Tauberian theorems techniques 
\cite{sei} one obtains for a $2^mNT_D$ graph:
\begin{equation}
\widetilde{d}_{2^m}=1 + {{\ln k} \over{\ln 2^m}}
\end{equation}
which represent the generalization of the result obtained for $m=1$.

\section{Random walks on $n NT_D$ and $p-polygon NT_D$ via asymptotic 
decimation}

The previous results can be extended to $nNT_D$,
where now $n$ is an integer and not necessarily a power of 2, and to 
$p\!-\!polygon NT_D$, where the branches of $NT_D$ 
are replaced by $p$-vertices regular polygons (Fig. 3). 
Let us consider $nNT_D$ first.
While relation (\ref{halley}) still holds, the exact time-rescaling procedure 
can not be applied to the branch of generic length $n$.
However even in this case it is possible to obtain an asymptotic recursion
relation applying the Renormalization Group techniques usually implemented 
on exactly decimable fractals \cite{nove}. Although this procedure cannot give
an exact equation for $\widetilde{P_O}^{tree}(\lambda)$ as in the previous case,
nevertheless it can be used to obtain the exact value of $\widetilde{d}$
via an asymptotic estimation.
Indeed, in this case the branch of the $nNT_D$ can be considered as
a tree with a dilatation factor equal to $n$. 
The log of this tree can be reduced to a unitary length log after the 
suppression of the $n-2$ sites between the edges and introducing a new link 
connecting the edges. The same operation can be repeated for branches of every 
length suppressing the inner $n-2$ consecutive sites in every sequence of $n$ 
sites and introducing a new link between the surviving points. The final 
structure is equal to the original tree and the generating function  
$\widetilde{P_A}^{branch}(\epsilon)$ becomes 
$\widetilde{P_A}^{'branch}(\epsilon')$ where:
\begin{equation}
\epsilon' = n^2 \epsilon
\label{tre}
\end{equation}
\begin{equation}
\widetilde{P_A}^{'branch}(\epsilon')= { 1\over n} 
\widetilde{P_A}^{branch}(\epsilon)
\label{quattro}
\end{equation}
Now $\widetilde{P_A}^{'branch}(\epsilon')$ coincides with
$\widetilde{P_O}^{tree}(\epsilon')$ since our branch has been transformed 
into a tree and (\ref{halley}) can be rewritten as:
\begin{equation}
\widetilde{P_O}^{tree}(\epsilon) ={{ n \widetilde{P_O}^{tree}(n^2 \epsilon) +k}
\over{2\epsilon
 n \widetilde{P_O}^{tree}(n^2 \epsilon) +k}}
\label{stella}
\end{equation}
Using the procedure described in the previous section for $2^m NT_D$, 
from (\ref{stella}) it follows that for an $n-NT_D$ the spectral dimension 
is given by:
\begin{equation}
\widetilde{d}_{n}=1 + {{\ln k} \over{\ln n}}
\end{equation}
An analogous technique can be used for $p\!-\!polygon NT_D$ (Fig. 3).
The log polygon has now $p$ faces of unitary length; from each of $p-1$ of its 
vertices $k$ polygons depart, whose faces have length $n$ and so on. These 
structures, though similar to $NT_D$ are no longer loopless structures nor 
necessarily bipartite graphs (e.g. the $3\!-\!polygon$ tree).
The Cutting-Decimation transform 
can be applied to $p\!-\!polygon NT_D$ as 
in the case of $NT_D$ with the same substitutions (\ref{tre}) and (\ref
{quattro}). Indeed, even if (\ref{halley}) does not hold
in this case, a new relation 
between the generating functions of the tree and that of one of its branches
can be obtained using bundled structures theory \cite{dieci}. 
A bundle structure is a 
composed graph obtained by joining to each point of a ``base graph" a copy of a 
``fibre" graph in such a way that every fibre has only one point in common with 
the base and no points in common with the other fibers. Let us consider a 
$p\!-\!polygon NT_D$ and suppose to attach 
$k$ branches also in the free vertex of the log
(the root of the tree): 
we obtain a bundled structure having the log polygon as base and 
the graph made of $k$ branches as fibre. 
Since for a $p$-polygon:
\begin{equation}
\widetilde{P_O}(\lambda) \sim {1 \over{ p (1- \lambda)}}
\end{equation}
as $\lambda \rightarrow 1$, we obtain for our bundled structure:
\begin{equation}
\widetilde{P_O}^{b.s.}(\lambda)={1 \over{1-{k\over{k+1}}\widetilde{F_A}
^{branch}(\lambda)}}{1\over p} \left(
1- { \lambda\over {k+1}}{1 \over { 1- 
{k\over{k+1}}\widetilde{F_A}^{branch}(\lambda)}} \right) ^{-1}
\label{primo}
\end{equation}
where $\widetilde{P_O}^{b.s.}(\lambda)$ is the generating function of the 
probability of returning to point $O$ (one of the vertices of the log polygon)
after a random walk on the bundled structure and $\widetilde{F_A}
^{branch}(\lambda)$ is the generating function of the probability of returning 
for the first time
to the point of connection with the base after a random walk on the fibre. 
For a generic structure the generating function of the probability of returning 
to the starting point $O$,  $\widetilde{P_O}(\lambda)$ and of returning for the 
first time to the starting point, $\widetilde{F_O}(\lambda)$, satisfy 
\cite{undici}:
\begin{equation}
\widetilde{P_O}(\lambda)= \left( 1- \widetilde{F_O}(\lambda) \right) ^{-1}
\label{secondo}
\end{equation}
Using:
\begin{equation}
\widetilde{F_O}^{b.s.}(\lambda)= {k \over {k+1}}
\widetilde{F_A}^{branch}(\lambda)+ {1\over {k+1}}\widetilde{F_O}^{tree}
(\lambda)
\label{terzo}
\end{equation}
where 
${F_O}^{tree}(\lambda)$ refers to the $p\!-\!polygon NT_D$, from (\ref{primo}), 
(\ref{secondo}) and (\ref{terzo}) a relation between 
$\widetilde{P_O}^{tree}(\lambda)$ and $\widetilde{P_A}^{branch}(\lambda)$
follows, which represents the cutting transformation.  
It is now possible to perform the Cutting-Decimation transform
to $p\!-\!polygon NT_D$ and get:
\begin{equation} 
\widetilde{d}_{p}=1 + {{\ln k(p-1)} \over{\ln n}}
\end{equation}
In the same way we can calculate the spectral dimension of an $NT_D$ built
with $d-$dimensional simplexes instead of $p\!-\!polygons$. A $d$-dimensional
simplex is a complete graph of $d+1$ points i.e. a graph where each point
is nearest neighbor of all other points. The $2-$dimensional case is the
triangle, the $3-$dimensional one is the tetrahedron and so on.
Since for $d-$simplex $\widetilde{P_O}(\lambda)\sim 1/(d+1)(1-\lambda)$
the spectral dimension is:
\begin{equation}
\widetilde{d}_{d}=1 + {{\ln kd } \over{\ln n}}
\end{equation}

\section{Conclusions}

The generalized $NT_D$ lattices described here have in general
non integer spectral dimension depending on the 
geometrical features of the lattices, such as the growing factor
$2^m$ or $n$, the number of branches $k$ and the number of vertices of the 
polygon. 
Since the intrinsic fractal dimension always coincides with the spectral 
dimension, the diffusion \cite{sei} 
on all these lattices is not anomalous i.e. is described by the asymptotic law:
\begin{equation}
<r^2(t)> \sim t^{\alpha}
\end{equation}
with $\alpha =1$. 
Moreover as in the case of 
simple $NT_D$ lattices \cite{otto}, 
we can show that also generalized $NT_D$ present 
dynamical dimension splitting. This means that the vibrational spectral 
dimension ${\overline d}$, which characterizes the density of vibrational 
modes as $\omega \rightarrow 0$ through the relation:
\begin{equation}
\rho(\omega) \sim \omega^{{\overline d}-1}
\end{equation}
is different from the diffusive spectral dimension of random walks 
$\widetilde{d}$. In particular we have ${\overline d}=1$ and this can be 
intuitively understood noting that the topology of generalized $NT_D$
is dominated by linear chains which become longer and longer 
in the outer branches. 
Since ${\overline d}=1$, by the generalized Mermin 
and Wagner Theorem \cite{dodici}, phase transitions 
with spontaneous breaking of a continuous symmetry for non zero temperature
cannot occur on these structures. 
All these remarkable properties, typical of simple $NT_D$ lattices, have been 
here obtained even in absence of some peculiar characteristics of $NT_D$ such
as the bipartite and loopless nature of the graph. This enlarged family of 
lattices can be used, as $NT_D$ lattices,
to reach a better understanding of the geometrical features giving rise
to dynamical dimension splitting and, at the same time, they represent 
the testing ground to study the relation between spectral dimensions and
physical phenomena.

\vspace{1cm}
\centerline{Figure captions}
\vspace{1cm}
\noindent {\bf Fig.1}  $2^mNT_D$ with $k=3$, $m=1$\\
    \\
{\bf Fig.2} Cutting-Decimation procedure:\\
a) Cutting of the log of the $NT_D$\\
b) Separation of the $k$ branches\\
c) Decimation of the points labelled by $X$\\
d) Recovering of the original $NT_D$\\
\\ 
{\bf Fig.3} $4-polygon NT_D$ with $k=1$, $n=2$

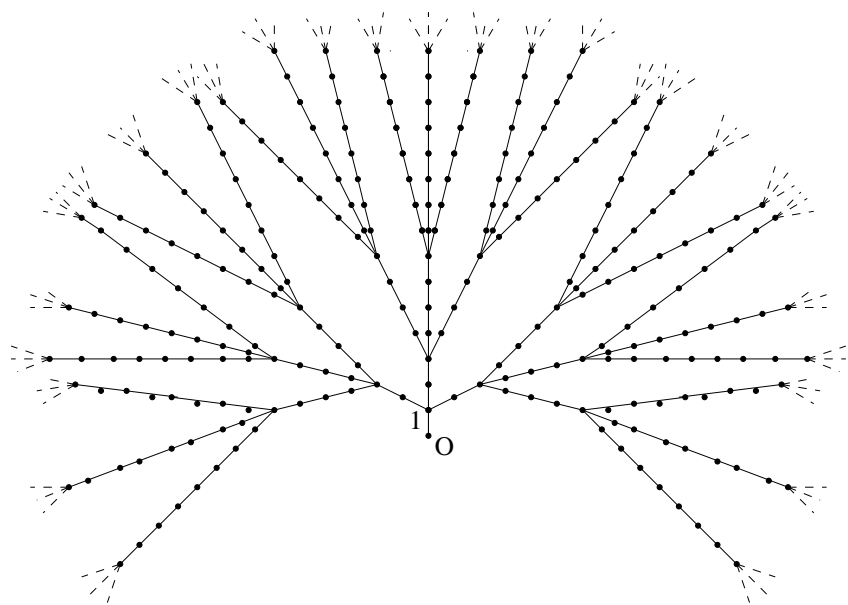
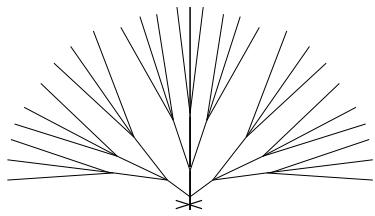
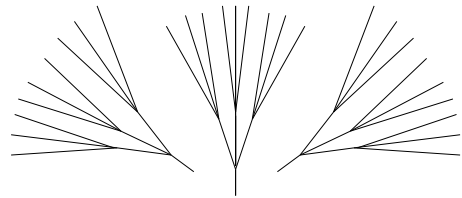
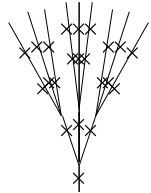
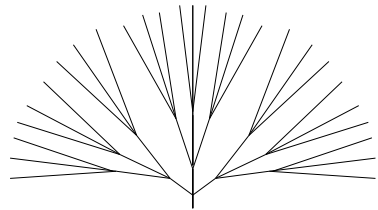
\end{document}